\documentclass[12pt,preprint]{aastex}
\usepackage{color}
\usepackage{graphics,graphicx}
\usepackage{rotating}

\newcommand{\lapprox} {\, \lower3pt\hbox{$\sim$}\llap{\raise2pt\hbox{$<$}}\,}
\newcommand{\gapprox} {\, \lower3pt\hbox{$\sim$}\llap{\raise2pt\hbox{$>$}}\,}

\begin{document}

\title{THE SPECIFIC ACCELERATION RATE IN LOOP-STRUCTURED SOLAR FLARES -- IMPLICATIONS FOR ELECTRON ACCELERATION MODELS}
\author{\sc
Jingnan~Guo\altaffilmark{1,3},
A.~Gordon~Emslie\altaffilmark{2},
AND Michele~Piana\altaffilmark{3,4}}

\altaffiltext{1}{Institut f\"{u}r Experimentelle und Angewandte Physik, Christian-Albrechts-Universit\"{a}t zu Kiel, 24118, Kiel, Germany; guo@physik-uni.kiel.de}

\altaffiltext{2}{Department of Physics and Astronomy, Western Kentucky University, Bowling Green, KY 42101; emslieg@wku.edu}

\altaffiltext{3}{Dipartimento di Matematica, Universit\`a di Genova, via Dodecaneso 35, 16146 Genova, Italy}

\altaffiltext{4}{CNR - SPIN, via Dodecaneso 33, I-16146 Genova, Italy; annamaria.massone@cnr.it; piana@dima.unige.it}

\begin{abstract}

We analyze electron flux maps based on {\em RHESSI\ } hard X-ray imaging spectroscopy data for a number of extended coronal loop flare events.  For each event, we determine the variation of the characteristic loop length $L$ with electron energy $E$, and we fit this observed behavior with models that incorporate an extended acceleration region and an exterior ``propagation'' region, and which may include collisional modification of the accelerated electron spectrum inside the acceleration region.  The models are characterized by two parameters: the plasma density $n$ in, and the longitudinal extent $L_0$ of, the acceleration region.  Determination of the best-fit values of these parameters permits inference of the volume that encompasses the acceleration region and of the total number of particles within it.  It is then straightforward to compute values for the emission filling factor and for the {\it specific acceleration rate} (electrons~s$^{-1}$ per ambient electron above a chosen reference energy).   For the 24 events studied, the range of inferred filling factors is consistent with a value of unity.  The inferred mean value of the specific acceleration rate above $E_0=20$~keV is $\sim$10$^{-2}$~s$^{-1}$, with a 1$\sigma$ spread of about a half-order-of-magnitude above and below this value.  We compare these values with the predictions of several models, including acceleration by large-scale, weak (sub-Dreicer) fields, by strong (super-Dreicer) electric fields in a reconnecting current sheet, and by stochastic acceleration processes.

\end{abstract}


\section{Introduction}\label{intro}

The Ramaty High Energy Solar Spectroscopic Imager \citep[{\em RHESSI},][]{linetal02} has revealed a new class of flares \citep{suetal04,vebr04, krucker08,xuetal08} in which the bulk of the hard X-ray emission is produced predominantly throughout the coronal portion of the loop, but for which the hard X-ray spectrum, and the variation of source size with energy \citep{xuetal08}, indicate strongly that such sources are {\it not} thermal.  Rather they involve the injection of nonthermal electrons into a coronal region which is not only the site of particle acceleration, but also dense enough to act as a thick target, stopping the accelerated electrons before they can penetrate to the chromosphere.  As discussed by \citet{xuetal08} and \citet{guoetal2012aa}, analysis of the variation of the longitudinal (field-aligned) source extent $L$ of such sources, as a function either of photon energy $\epsilon$ (deduced from images of the hard X-ray flux) or of electron energy $E$ \citep[deduced from electron flux maps --][]{pianaetal07} can be used to obtain information on the size of the acceleration region and the density in the region into which the accelerated electrons propagate.  Further, as discussed in \citet{guoetal2012}, this information, coupled with information obtained through parametric fitting of the spatially-integrated hard X-ray spectrum, can be used to obtain information on the filling factor (the ratio of hard-X-ray-emitting volume to the total volume encompassing the source(s)) and on the {\it specific acceleration rate}, the ratio of the rate (s$^{-1}$) of acceleration of electrons above a chosen reference energy to the number of electrons available for acceleration.

While it is clear that collisional modification to the accelerated electrons occurs in the ``propagation'' region exterior to the acceleration region, an interesting question is: ``should collisional modification to the accelerated electron spectrum inside the acceleration region itself also be considered?'' On the one hand, the presence of a finite ambient density in the acceleration region should require inclusion of such a term.  On the other hand, it could be argued either (1) that the acceleration mechanism ``swamps'' any effects of Coulomb collisions, or (2) that collisional processes are already included in the physics governing the form of the accelerated electron spectrum, so that an additional ``post-acceleration'' collisional term need not be considered.  In either of these latter cases, we should model the acceleration region by an effectively ``tenuous'' region which has zero density {\it for the purposes of calculating post-acceleration collisional modifications to the electron spectrum}, although it must be stressed that the actual density in the acceleration region is likely to be comparable to that in the exterior ``propagation'' region, so that substantial bremsstrahlung emission will still be produced in the acceleration region.

Although ``tenuous'' acceleration region models have been considered by a number of authors \citep{emslie2008AIP,xuetal08,2011ApJ...730L..22K,guoetal2012aa,guoetal2012}, so far only \citet{xuetal08} have considered the ``dense'' acceleration regiom model.  Here we extend the results of \citet{guoetal2012} to include both ``tenuous'' and ``dense'' acceleration region scenarios.

In Section~\ref{events} we present the list of 24 extended-loop events studied.  In Section~\ref{analysis} we derive analytical forms for $L(E)$, both for the ``tenuous'' acceleration region model and for two different ``dense'' acceleration region models, one with a spatially uniform injection term over a finite length and one with a Gaussian spatial profile of the injection term.  Fitting all three of these forms to the observationally-inferred forms of $L(E)$ results in corresponding best-fit estimates of the acceleration region length $L_0$ and density $n$.  These are then used (Section~\ref{derived_quantities}) to infer, for each event, the value of the emission filling factor $f$ (the fraction of the observed volume in which hard X-ray emission occurs).  This is in turn used to estimate the {\it acceleration region filling factor} $f_{\rm acc}$ (the ratio of the volume that is actively involved in electron acceleration to the overall volume that encompasses the acceleration region(s)), and so to obtain values of the specific acceleration rate $\eta(\ge E_0)$, the rate of acceleration of electrons to energies $\ge E_0$ divided by the number of ambient electrons available for acceleration.  In Section~\ref{result} we summarize the salient features of the results obtained and in Section~\ref{discussion} we compare the ensemble-averaged values of the quantity $\eta(\ge E_0)$ with the predictions of various acceleration models.  In Section~\ref{conclusions} we summarize the results obtained and present our conclusions.

\section{Observations}\label{events}

\begin{deluxetable} {cccccccc}
    \tablewidth{0pt}
    \tabletypesize{\scriptsize}
    \tablecaption{Event List and Spectral Fit Parameters\label{table:spec}}
    \tablehead{
    \colhead{Event No.} & \colhead{Date} & \colhead{Time (UT)} & \colhead{EM ($10^{49}$~cm$^{-3}$)} & \colhead{T (keV) } & \colhead{$\delta$} & \colhead{$E_t$ (keV)} & \colhead{$d{\cal N}/dt$ ($10^{35}$~s$^{-1}$) } }
    \startdata
	 1 & 2002-04-12 & 17:42:00-17:44:32 & $0.30$ & $1.53$ & $8.24$ & $15.5$ & $2.71$ \\
	 2 & 		   & 17:45:32-17:48:00 & $0.46$ & $1.54$ & $8.01$ & $15.5$ & $4.69$ \\
	  \hline
 	3 & 2002-04-15 & 00:00:00-00:05:00 & $0.22$ & $1.75$ & $7.48$ & $15.5$ & $4.70$ \\
	 4 & 		   & 00:05:00-00:10:00 & $0.76$ & $1.61$ & $7.93$ & $15.5$ & $9.32$ \\
	 5 & 		   & 00:10:00-00:15:00 & $1.02$ & $1.60$ & $8.37$ &$15.5$ & $11.41$ \\
	  \hline
 	6 & 2002-04-17 & 16:54:00-16:56:00 & $0.06$ & $1.51$ & $5.70$ & $15.5$ & $0.39$ \\
	 7 & 		   & 16:56:00-16:58:00 & $0.22$ & $1.43$ & $8.78$ & $14.8$ & $2.43$ \\
	  \hline
 	8 & 2003-06-17 & 22:46:00-22:48:00 & $1.92$ & $1.71$ & $9.95$ &$16.5$ & $17.27$ \\
	 9 & 		   & 22:48:00-22:50:00 & $2.59$ & $1.67$ & $10.36$ &$16.5$ & $17.91$ \\
	  \hline
 	10 & 2003-07-10 & 14:14:00-14:16:00 & $1.26$ & $1.45$ & $10.05$ & $15.5$ & $7.43$ \\
	 11 & 		   & 14:16:00-14:18:00 & $1.31$ & $1.34$ & $10.38$ & $14.8$ & $8.53$ \\
	  \hline
 	12 & 2004-05-21 & 23:47:00-23:50:00 & $0.35$ & $1.85$ & $7.07$ & $18.5$ & $3.28$ \\
	 13 & 		   & 23:50:00-23:53:00 & $0.62$ & $1.75$ & $7.51$ & $18.5$ & $2.32$ \\
	 \hline
 	14 & 2004-08-31 & 05:31:00-05:33:00 & $0.06$ & $1.61$ & $10.56$ & $15.5$ & $0.40$ \\
	 15 & 		   & 05:33:00-05:35:00 & $0.21$ & $1.57$ & $12.19$ & $18.5$ & $0.29$ \\
	 16 & 		   & 05:35:00-05:37:00 & $0.29$ & $1.48$ & $7.45$  & $18.5$ & $0.16$ \\
	  \hline
 	17 & 2005-06-01 & 02:40:20-02:42:00 & $0.14$ & $1.81$ & $6.53$ & $17.5$ & $1.44$ \\
	18  & 		   & 02:42:00-02:44:00 & $0.37$ & $1.70$ & $7.86$ & $17.5$ & $2.67$ \\
	  \hline
 	19 & 2011-02-13 & 17:33:00-17:34:00 & $0.54$ & $1.39$ & $5.86$ & $10.5$ & $35.02$ \\
	 20 & 		   & 17:34:00-17:35:00 & $0.52$ & $1.68$ & $6.55$ & $14.5$ & $19.43$ \\
	  \hline
 	21 & 2011-08-03 & 04:31:12-04:33:00 & $0.36$ & $1.61$ & $9.23$ & $15.5$ & $3.96$ \\
 	\hline
 	22 & 2011-09-25 & 03:30:36-03:32:00 & $0.13$ & $1.44$ & $8.33$ & $14.5$ & $1.19$ \\
    \hline
	23 & 2005-08-23 & 14:23:00-14:27:00 & $0.04$ & $2.12$ & $6.32$ & $14.2$ & $5.44$ \\
	24 &            & 14:27:00-14:31:00 & $0.29$ & $1.87$ & $7.96$ & $16.5$ & $10.20$\\
 	\hline
\enddata
\end{deluxetable}

The list of events\footnote{In this context, an ``event'' is a time interval during a flare for which spatial and spectral observations are sufficiently good to permit both a determination of the source spatial structure at a variety of energies and the overall spectrum of the hard X-ray emission.  As can be seen in Table~\ref{table:spec}, some flares provide multiple ``events''; other flares only one.} studied is shown in Table~\ref{table:spec}; these are the same as those used by \citet{guoetal2012}, with the addition of two additional events associated with a flare on 2005~August~23 (Event \#s 23 and 24). For each event, we fit the spatially-integrated hard X-ray emission spectrum with an isothermal-plus-power-law form, yielding values of the emission measure EM (cm$^{-3}$) and temperature $T$ (keV) of the thermal source, the intensity and spectral index $\delta=\gamma+1$ of the injected nonthermal electron spectrum (corresponding to the hard X-ray spectral index $\gamma$), and $E_t$ (keV), the transition energy between the thermal and nonthermal components of the hard X-ray spectrum.  Straightforward thick-target modeling \citep[e.g.,][]{brown71} then provides values of $d{\cal N}/dt(\ge E_0)$ (s$^{-1}$), the rate of acceleration of electrons to energies above the (arbitrary) reference energy $E_0 = 20$~keV.

For each event, we also created electron flux images at a series of electron energies $E$, each produced using electron visibilities constructed via the procedure of \citet{pianaetal07} and the uv$\_$smooth image reconstruction algorithm \citep{massone2009}.  At each energy $E$, the (field-aligned) root-mean-square length $L$ and (cross-field) width $W$ of the source were then calculated using the procedure of \citet{guoetal2012aa}.

\section{Analysis}\label{analysis}

\subsection{Tenuous Extended Acceleration Region Model}\label{tenuous_model}

\citet{guoetal2012} fit the observed form of $L(E)$ to a ``tenuous'' acceleration region model \citep{xuetal08}. In such a model, either (a) the acceleration process is considered to dominate the evolution of the electron spectrum to such an extent that the effect of collisions in the acceleration region can be ignored, or (b) the effect of collisions is already incorporated in the production of the accelerated electron spectrum.  In either case, no additional collisional modification to the electron spectrum need be considered.  Thus, the ``tenuous'' acceleration region is considered to have zero density  {\it only in respect of the variation of the electron spectrum throughout the acceleration region}.  However, it must again be stressed that in such a model there is still a finite density (and so a bremsstrahlung target) in the acceleration region, so that the hard X-ray source includes the acceleration region itself.

Consider, then, electrons with an injected spectrum $F_0(E_0) \sim E_0^{-\delta}$ (cm$^{-2}$~s$^{-1}$~keV$^{-1}$) that are accelerated within a region extending over [$-L_{0,t}/2$,$L_{0,t}/2$] and propagate through an exterior region with uniform density $n_t$, in which the electrons suffer energy loss through Coulomb collisions with ambient particles, particularly electrons \citep{1978ApJ...224..241E}.  Since collisions in the acceleration region are neglected, the calculated value $L_t(E)$ in such a model consists of two parts: the (energy-independent) length of the acceleration region $L_{0,t}$ and the (injection-spectrum-weighted) average penetration depth associated with the electrons of injection energies $E_0 \ge E$ that contribute to the electron map at energy $E$. For Coulomb collisions in a uniform ambient target, the latter term is equal to a spectral shape factor times the quantity $E^2/2Kn_t$, where $K = 2 \pi e^4 \Lambda$, $e$ being the electronic charge and $\Lambda$ the Coulomb logarithm. Quantitatively, $L_t(E)$ has the simple quadratic form \citep[see][]{guoetal2012}:

\begin{equation}\label{model-electron-tenuous}
L_t(E)= L_{0,t} + \frac{2}{Kn_t} \, \sqrt{\frac{2}{(\delta - 3)(\delta-5)}} \, \, E^2 \, .
\end{equation}
Least-squares fitting of the observed values of $L(E)$ to this analytic form leads to best-fit values of the acceleration region length $L_{0,t}$ and the density $n_t$ in the exterior propagation region, which we also take to be the density in the acceleration region itself.  The resulting values of $L_{0,t}$ and $n_t$ have been given in \citet{guoetal2012}, and, for completeness, these results are repeated here in Table~2.

\subsection{Dense Extended Acceleration Region Model}\label{dense_model}

As noted above, the form~(\ref{model-electron-tenuous}) does not include the effects of collisional modification to the accelerated electron spectrum within the acceleration region itself. Therefore, for completeness we here, following \citet{xuetal08}, develop the expression for $L(E)$ associated with a ``dense'' acceleration region model, which {\it does} take into account post-acceleration collisional losses within the acceleration region itself.  For the tenuous acceleration model, the lack of collisional modification to the electrons inside the acceleration region means that the form for $L_t(E)$ -- Equation~(\ref{model-electron-tenuous}) -- does not depend on the form of the variation of the electron injection rate with position in the acceleration region.  However, for the dense acceleration model, collisional modification to the electron spectrum in the acceleration region means that the results {\it do} depend on the spatial form of the acceleration rate.  Hence, we have evaluated the form of $L(E)$ for two illustrative parametric forms of the spatial profile of the accelerated electron flux, in order to evaluate the sensitivity of the results to the particular form of this spatial profile.

\subsubsection{Spatially Uniform Injection Profile}\label{uniform_inj}

We first consider an acceleration region extending over [$-L_{0,u}/2$,$L_{0,u}/2$], with a uniform density $n_u$ equal to that in the rest of the loop and standard collisional losses appropriate to such a density.  Electrons, with a spectrum $F_0(E_0) \sim E_0^{-\delta}$, are assumed to be injected uniformly throughout this region, and the electrons injected at each location $s_0$ ($-L_{0,u}/2 \leq s_0 \leq L_{0,u}/2$) propagate in both directions along the loop. The electron flux spectrum at longitudinal position $s$ then follows from the one-dimensional continuity equation:

\begin{equation}\label{fes_dense_rec}
F_u(E,s, s_0) = F_0(E_0) \, {dE_0 \over dE} \sim {E \over (E^2 + 2 K n_u |s - s_0|)^{(\delta + 1)/2}} \,\,\, ,
\end{equation}
where we have used the collisional relation $E_0^2 = E^2 + 2 KN = E^2 + 2 K n_u |s - s_o|$, $N$ being the column density (cm$^{-2}$) along the direction of electron propagation.  The corresponding expression for the longitudinal loop extent $L(E)$ is:

\begin{equation}\label{model-dense_rec}
L_u(E) = \sqrt{\int_{-\infty}^\infty \int_{-L_{0,u}/2}^{L_{0,u}/2} s^2 \, F_u(E,s, s_0) \, ds_0 \, ds \over \int_{-\infty}^\infty \int_{-L_{0,u}/2}^{L_{0,u}/2} F_u(E,s, s_0) \, ds_0 \, ds} \,\,\, .
\end{equation}
Note that we have formally approximated the limits on the $s$-integration at $\pm \infty$, rather than using an estimated overall loop extent $\pm L_{\rm max}$.  This approximation was used principally to avoid having to estimate the total loop length, including the portion occupied by weakly-emitting regions near the chromosphere which, especially for events far from the limb, are the regions where the observed geometry is most sensitive to projection effects.  However, since $L_u(E)$ is equal to the square root of a {\it ratio} of integrals, the results do not depend significantly on the values of the $s$-integration limits used.

The forms~(\ref{fes_dense_rec}) and~(\ref{model-dense_rec}) do not result in a closed analytic form for $L_u(E)$ (cf. Equation~(\ref{model-electron-tenuous})) and so a straightforward least-squares fitting procedure cannot be used to determine the best-fit values of the acceleration region length $L_{0,u}$ and density $n_u$.  However, these best-fit values can nevertheless be determined numerically by exploring $(L_{0,u},n_u)$ space and computing, at each sampled point, the value of $\chi^2_u (\equiv \sum^{E_{max}}_{E_{min}} w(E)[L_u(E)-L(E)]^2 /n_{free})$, where the point weightings\footnote{In order to provide quantitative uncertainties $L_{err}(E)$ on observed loop lengths $L(E)$, we applied a Monte Carlo approach in which random noise is added to the visibilities and the resulting uv-smooth images recomputed and reanalyzed \citep{guoetal2012aa}.} $w(E) = 1/L_{err}^2(E)$ and $n_{free}$ is the number of degrees of freedom for the fitting.  The best-fit values of $L_{0,u}$ and $n_u$ are determined by finding the location of the minimum value of $\chi^2_u$.

\begin{center}
\begin{figure}[pht]
\begin{tabular}{cc}
{ \includegraphics[width=0.45\textwidth]{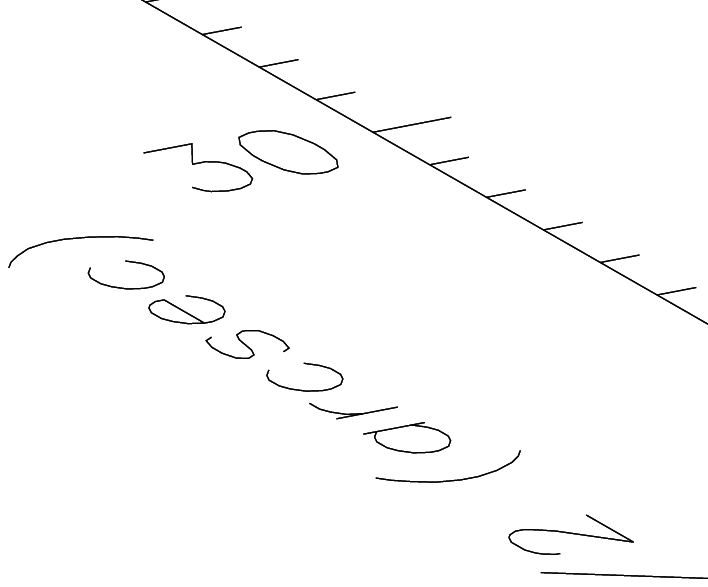} }
{ \includegraphics[width=0.45\textwidth]{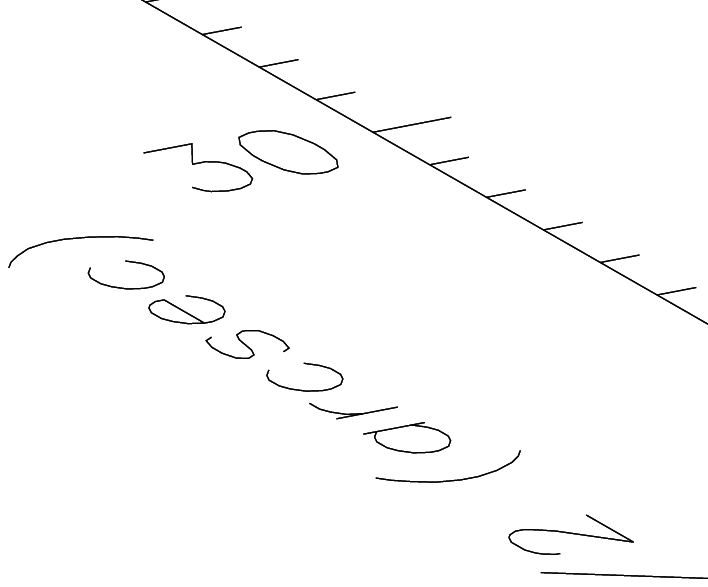} }
\end{tabular}
\caption{\textit{Left}: The $\chi^2_u(L_{0,u},n_u)$ surface for Event \#1 (2002~April~12).  \textit{Right}: The $\chi^2_u(L_{0,u},n_u)$ surface for Event \#9 (2003~June~17).  The minimum values of $\chi^2_u$ are also shown.}
\label{fig_uniform_case}
\end{figure}
\end{center}

The left panel of Figure~\ref{fig_uniform_case} shows the $\chi^2_u(L_{0,u},n_u)$ surface for Event \#1 (2002~April~12), together with contours of $\chi^2_u$ equal to 1, 2, 5, and 10. A pronounced local minimum ($\chi^2_u = 0.13$) at ($L_{0,u} = 33$~arcseconds, $n_u = 1.2 \times 10^{11}$~cm$^{-3}$) is found.  By contrast, the right panel of Figure~\ref{fig_uniform_case} shows the $\chi^2_u(L_0,n_u)$ surface for Event \#9 (2003~June~17), together with contours of $\chi^2_u$ equal to 3, 5, and 10.  For this event, the local minimum of $\chi^2_u$, and so the best-fit values ($L_{0,u} = 35$~arcseconds, $n_u=1.0 \times 10^{11}$~cm$^{-3}$) are, especially for the density $n_u$, not so precisely determined, and this is reflected in the associated higher value $\chi^2_u=2.07$ in Table~2.

The best-fit values of of $L_{0,u}$ and $n_u$, and the corresponding values of $\chi^2_u$, for all 24 events are listed in Table~2.

\subsubsection{Gaussian Injection Profile}\label{gaussian_inj}

Here we again assume an extended, dense acceleration region, but for which the spatial distribution of the energetic electron injection falls off with distance $s_0$ from the ``kernel'' of the acceleration site (of uniform density $n_G$) according to a Gaussian form characterized by a standard deviation $\sigma_0$.  For such an injection profile, the form of the electron flux spectrum $F_G(E,s, s_0)$ is

\begin{equation}\label{fes_dense_gauss}
F_G(E,s, s_0) \sim  {\frac{\exp(-{s_0}^2/2{\sigma_0}^2)}{\sigma_0}} {E \over {(E^2 + 2 K n_G |s - s_0|)^{(\delta + 1)/2}}} \, ,
\end{equation}
which is substituted into

\begin{equation}\label{model-dense_gauss}
L_G(E) = \sqrt{\int_{-\infty}^\infty \int_{-\infty}^\infty s^2 \, F_G(E,s, s_0) \, ds_0 \, ds \over \int_{-\infty}^\infty \int_{-\infty}^\infty  F_G(E,s, s_0) \, ds_0 \, ds} \,\,\,.
\end{equation}
to determine the form of the longitudinal loop extent $L_G(E)$.  The best-fit values for $\sigma_0$ and $n_G$ can then be derived by a procedure similar to that for the uniform injection case.

\begin{center}
\begin{figure}[pht]
\begin{tabular}{cc}
{ \includegraphics[width=0.45\textwidth]{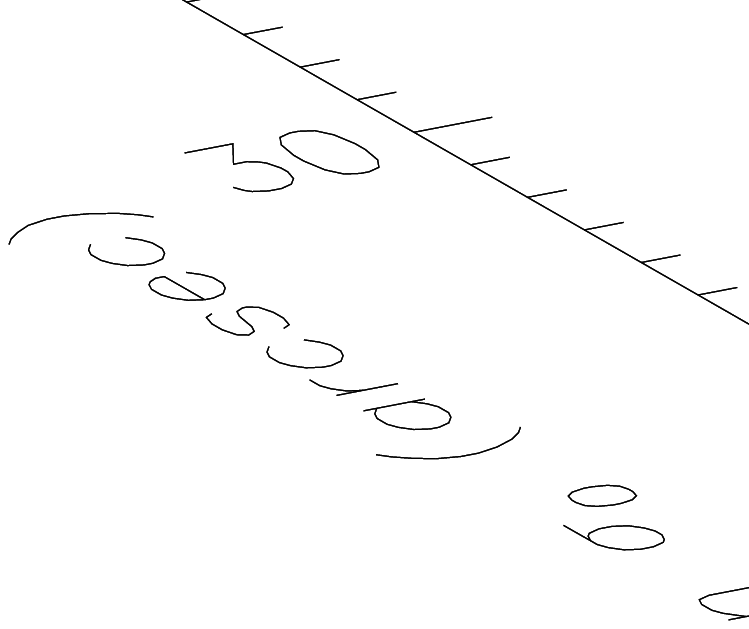} }
{ \includegraphics[width=0.45\textwidth]{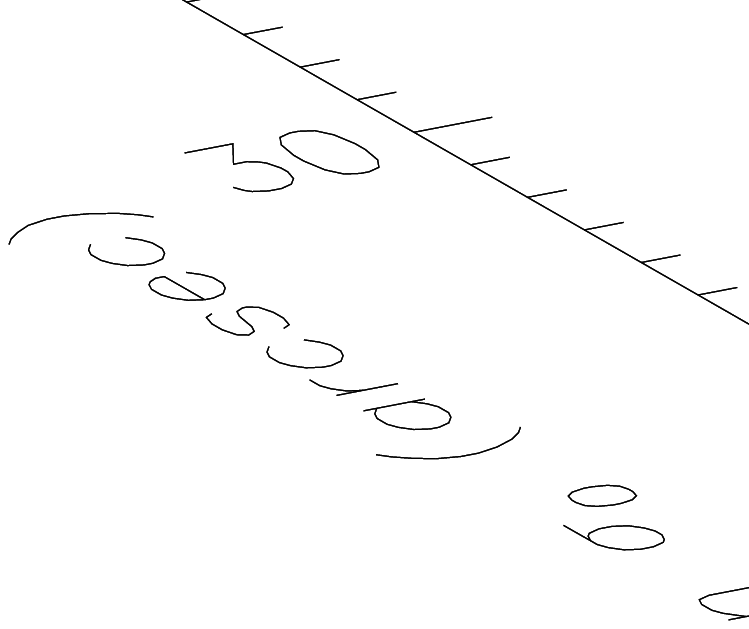} }
\end{tabular}
\caption{\textit{Left}: The $\chi^2_G(\sigma_0,n_G)$ surface obtained for Event \#1 (2002~April~12). \textit{Right}: The $\chi^2_G(\sigma_0,n_G)$ surface for Event \#9 (2003~June~17).  The minimum values of $\chi^2_G$ are also shown.}
\label{fig_gaussian_case}
\end{figure}
\end{center}

The left panel of Figure~\ref{fig_gaussian_case} shows the $\chi^2_G(\sigma_0,n_G)$ surface for Event \#1 (2002~April~12), together with contours of $\chi^2_G$ equal to 1, 2, 5 and 10. A pronounced local minimum ($\chi^2_G=0.12$) at ($\sigma_0 = 13.5$~arcseconds, $n_G = 1.2 \times 10^{11}$~cm$^{-3}$) is found.  By contrast, the right panel of Figure~\ref{fig_gaussian_case} shows the $\chi^2_G(\sigma_0,n_G)$ surface for Event \#9 (2003~June~17).  Again, for the latter event, the local minimum of $\chi^2_G$ (and so the best-fit values of $\sigma_0 = 17.5$~arcseconds and $n_G=0.9 \times 10^{11}$~cm$^{-3}$) are not so precisely determined.  This is
especially true for the value of the characteristic scale $\sigma_0$, and this lack of precision is reflected in the associated high value $\chi^2_G=2.05$ in Table~2.

We define the characteristic length of the acceleration region as the full width of the Gaussian function at half the maximum (FWHM) $L_{0,G}=2 \, \sqrt{2 \, {\rm{ln}}2} \, \sigma_0$. The best-fit results for $L_{0,G}$ and $n_G$, and the corresponding values of $\chi^2_G$, for all 24 events are shown in Table~2.

In general, the values of both $L_0$ and $n$ for the Gaussian injection profile are comparable to those for the uniform injection case, illustrating that the results are not critically dependent on the spatial profile of the acceleration rate within the acceleration region.

\section{Derived Quantities}\label{derived_quantities}

\subsection{Emission Filling Factor}\label{filling}

The soft X-ray emission measure EM is related to the plasma density $n$ and the {\it emitting} volume $V_{\rm emit}$ through EM $= n^2 \, V_{\rm emit}$. Given that an emitting region (considered here to be a cylinder of length $L$ and width $W$) may be composed of a number of discrete emitting subregions (e.g., ``strands,'' ``kernels''), the emitting volume may be equal to or smaller than the total flare volume

\begin{equation}\label{volume}
V = {\pi \over 4} \, W^2 \, L(E_t) \, ,
\end{equation}
a quantity readily ascertained from {\em RHESSI} images at the transition energy $E_t$ (Table~\ref{table:spec}), the maximum energy at which thermal emission is predominant.

The ratio of the emitting volume $V_{\rm emit}$, determined from the values of the emission measure EM (Table~\ref{table:spec}) and the inferred acceleration region density ($n_u$ or $n_G$, as appropriate; see Table~2), to the volume $V$ of the observed region that encompasses the emitting region(s) is termed the {\it emission filling factor} (or simply ``filling factor''):

\begin{equation}\label{filling-factor}
f = {V_{\rm emit} \over V} = \frac{{\rm EM}}{n^2 V} \,\,\, .
\end{equation}
Values of $f$ for each event, and the geometric mean and multiplicative 1$\sigma$ uncertainty for the ensemble of events studied, are given in Table~2.

\subsection{Specific Acceleration Rate}\label{eta_section}

The {\it specific acceleration rate} $\eta(\ge E_0)$ (electrons~s$^{-1}$ per ambient electron) is defined \citep{emslie2008AIP} as the ratio of two quantities: $d{\cal N}/dt (\ge E_0)$, the rate of acceleration of electrons to energies greater than $E_0$, and ${\cal N}$, the number of particles available for acceleration:

\begin{equation}\label{acceleration-rate}
\eta(\ge \!\! E_0) = \frac{1}{\cal{N}} \, \frac{d \cal{N}}{dt} (\ge E_0) \,\,\, .
\end{equation}
The quantity $d{\cal N}/dt (\ge$20~keV) is readily determined by spectral fitting of the spatially-integrated hard X-ray emission; values for all 24 events are given in Table~\ref{table:spec}.  The quantity ${\cal N}$ can be estimated from the relation

\begin{equation}\label{calN}
{\cal N} = f_{\rm acc} \, {\pi W^2 \over 4} \, L_0 \, n \, ,
\end{equation}
where the values of the acceleration region length $L_0$ and density $n$ are inferred from the appropriate acceleration region model and $f_{\rm acc}$ is the {\it acceleration region filling factor},\footnote{Note that the appearance of the $f_{\rm acc}$ term in the expression for ${\cal N}$ was implicitly assumed to be unity in the earlier work of \citet{guoetal2012}.  Here we elaborate on this assumption further.} the ratio of the volume in which acceleration actually occurs to the encompassing volume of the region that contains the acceleration site(s).  This quantity is not to be confused with the emission filling factor $f$, since the regions in which acceleration occurs are not necessarily congruent to the regions that produce hard X-ray emission.

We have no direct knowledge of the value of the acceleration region filling factor $f_{\rm acc}$.  However, for the coronal-loop sources considered here, it appears reasonable to assume that the volume in which acceleration occurs is a strict subset of the acceleration-plus-propagation volume.  Since hard X-ray emission occurs in both the acceleration and propagation regions, it follows that the acceleration filling factor is less than the emission filling factor, i.e., that $f_{\rm acc}< f$.  Further, a value $f_{\rm acc}$ that is less than unity reduces the number of electrons ${\cal N}$ available for acceleration, and therefore, for a given acceleration rate $d{\cal N}/dt(\ge E_0)$, the specific acceleration rate $\eta(\ge E_0)$ is increased.  As we shall see, the values of $\eta(\ge E_0)$  obtained for $f_{\rm acc}=1$ already impose significant constraints on particle acceleration models, so that a value $f_{\rm acc} \ll 1$ is extremely unlikely.

Table~2 shows that the values of the emission filling factor $f$ do vary considerably from event to event, and indeed that the values of $f$ obtained for some events are greater than one (and thus unphysical). Such results are doubtless artifacts of vagaries\footnote{For example, the derivation of quantitative electron flux maps depends \citep{pianaetal07} on the form of the bremsstrahlung cross-section used.  This cross-section
can vary a great deal with the degree of directivity of the exciting electron beam \citep{2004ApJ...613.1233M}.} in the inferred electron maps and/or in the modeling assumptions used.  Apart from a few unusual cases (discussed in Section~\ref{result}), the value of the emission filling factor $f$ is generally bounded above by $\sim$(2-3), well within the bounds of possible uncertainty in the quantitative electron flux values inferred from the {\em RHESSI} data and in the modeling assumptions used.  Moreover, the geometric mean of the emission filling factor $f$ for the ensemble of 24 events studied is very close to unity, a value hitherto tacitly assumed by many authors. We therefore assume that $f=1$ (i.e., that the entire observed region $V$ participates in both the hard X-ray emission process) and, based on the arguments in the previous paragraph, we further make the rather bold assumption that the emission and acceleration regions are congruent, so that $f_{\rm acc}=1$.  In doing so, we recognize that this may be an overly constraining assumption for models in which the acceleration region is intrinsically a small fraction of the region in which the electrons propagate and emit bremsstrahlung (see, e.g., Section~\ref{super_dreicer}).

From Equations~(\ref{acceleration-rate}) and~(\ref{calN}), with $f_{\rm acc}=1$, we can deduce the value of the specific acceleration rate $\eta (E_0)$ in each event; values (for $E_0 = 20$~keV), together with the geometric mean for the ensemble and its 1$\sigma$ multiplicative uncertainty, are given in Table~2.

\begin{deluxetable} {c|ccc|ccc|ccc|ccc|ccc|ccc}
    \tablewidth{0pt}
    \tabletypesize{\scriptsize}
    \tablecaption{Acceleration Region Characteristics}\label{table:params}
    \tablehead{ \colhead{Event No.} & \colhead{$L_{0,t}$} & \colhead{$L_{0,u}$} & \colhead{$L_{0,G}$} & \colhead{$n_t$} & \colhead{$n_u$} & \colhead{$n_G$} & \colhead{$f_t$} & \colhead{$f_u$} & \colhead{$f_G$} & \colhead{$\eta_t$} & \colhead{$\eta_u$} & \colhead{$\eta_G$} & \colhead{${\chi^2}_t$} & \colhead{${\chi^2}_u$} & \colhead{${\chi^2}_G$} \\
    & & $\!\!\!\!\!\!\!\!\!\!\!$ (arcsec) $\!\!\!\!\!\!\!\!\!\!\!$ & & & $\!\!\!\!\!\!\!\!\!\!\!$($10^{11}$~cm$^{-3}$)$\!\!\!\!\!\!\!\!\!\!$ & & & & & & $\!\!\!\!\!\!\!\!\!\!\!\!\!\!\!\!\!\!\!\!\!\!\!\!$ $(\ge 20$~keV; $\times 10^{-3}$~s$^{-1}$) $\!\!\!\!\!\!\!\!\!\!\!\!\!\!\!\!\!\!\!\!\!\!\!\!$ & & & & }
   \startdata
   1 & 18.6 & 33.0 & 31.8 & 1.5 & 1.2 & 1.2 & 0.45 & 0.69 & 0.69 & 6.5 & 4.6 & 4.8 & 0.12 & 0.13 & 0.12 \\
   2 & 16.3 & 31.0 & 24.7 & 1.4 & 1.4 & 1.3 & 0.83 & 0.82 & 1.08 & 14.5 & 7.3 & 10.5 & 0.33 & 0.37 & 0.33 \\
   \hline
   3 & 16.7 & 29.0 & 24.7 & 4.4 & 1.1 & 1.3 & 0.04 & 0.56 & 0.47 & 4.0 & 8.9 & 9.5 & 0.31 & 0.34 & 0.26 \\
   4 & 16.6 & 31.0 & 24.7 & 4.8 & 1.7 & 1.5 & 0.11 & 0.93 & 1.12 & 7.3 & 11.4 & 15.6 & 0.34 & 0.41 & 0.28 \\
   5 & 16.6 & 29.0 & 22.4 & 10.5& 2.6 & 2.3 & 0.03 & 0.43 & 0.56 & 3.3 & 7.5 & 11.1 & 0.41 & 0.52 & 0.82 \\
   \hline
   6 & 11.9 & 19.0 & 13.0 & 4.9 & 1.0 & 0.9 & 0.02 & 0.39 & 0.43 & 0.6 &2.1 &3.2 & 0.14 & 0.34 & 0.36 \\
   7 & 10.4 & 19.0 & 15.3 & 1.8 & 0.8 & 0.9 & 0.44 & 2.41 & 1.83 & 12.1 & 15.5 & 16.7 & 0.18 & 0.24 & 0.26 \\
   \hline
   8 & 17.8 & 33.0 & 29.4 & 2.6 & 0.8 & 0.9 & 0.90 & 8.99 & 8.20 & 24.1 & 41.9 & 34.0 & 1.10 & 2.70 & 2.70 \\
   9 & 18.8 & 35.0 & 41.2 & 2.9 & 1.0 & 0.9 & 1.05 & 8.69 & 10.45 & 23.1 & 35.5 & 19.2 & 1.78 & 2.07 & 2.05 \\
   \hline
   10 & 15.1 & 27.0 & 22.4 & 2.9 & 0.8 & 1.0 & 0.72 & 10.47 & 6.03 & 13.8 & 32.5 & 34.1 & 0.10 & 0.54 & 0.28 \\
   11 & 16.0 & 29.0 & 22.4 & 1.9 & 0.6 & 0.7 & 1.95 & 17.35 & 15.83 & 27.8 & 46.8 & 63.6 & 0.52 & 0.47 & 0.44 \\
   \hline
   12 & 10.3 & 19.0 & 13.0 & 5.1 & 1.9 & 1.7 & 0.08 & 0.54 & 0.65 & 4.9 & 7.1 & 11.4 & 0.80 & 0.75 & 0.95 \\
   13 & 9.9  & 19.0 & 13.0 & 4.6 & 1.9 & 1.7 & 0.18 & 1.02 & 1.23 & 4.1 & 5.1 & 8.2  & 0.93 & 0.92 & 0.99 \\
   \hline
   14 & 21.5 & 37.0 & 29.4 & 1.5 & 0.4 & 0.4 & 0.13 & 2.06 & 1.88 & 1.4 & 3.3 & 1.5 & 0.05 & 0.06 & 0.05 \\
   15 & 17.4 & 31.0 & 24.7 & 0.8 & 0.3 & 0.3 & 1.03 & 9.29 & 10.19 & 1.7 & 2.9 & 3.8 & 0.09 & 0.12 & 0.12 \\
   16 & 17.8 & 29.0 & 20.0 & 2.3 & 0.5 & 0.5 & 0.18 & 3.44 & 4.14 & 0.3 & 0.9 & 1.4 & 0.25 & 0.15 & 0.16 \\
   \hline
   17 & 11.0 & 21.0 & 15.3 & 3.9 & 1.4 & 1.4 & 0.05 & 0.39 & 0.39 & 2.9 & 4.3 & 5.9 & 0.17 & 0.28 & 0.30 \\
   18 & 9.9  & 19.0 & 12.9 & 3.2 & 1.4 & 1.3 & 0.22 & 1.21 & 1.45 & 7.0 & 8.7 & 14.0 & 0.38 & 0.41 & 0.55 \\
   \hline
   19 & 19.9 & 35.0 & 36.5 & 11.1 & 1.8 & 1.7 & 0.02 & 0.67 & 0.74 & 13.6 & 47.3 & 47.5 & 0.82 & 0.69 & 0.67 \\
   20 & 14.5 & 25.0 & 17.7 & 5.2  & 1.3 & 1.3 & 0.10 & 1.58 & 1.74 & 23.4 & 53.5 & 79.3 & 0.15 & 0.21 & 0.22 \\
   \hline
   21 & 9.9 & 19.0 & 12.9  & 2.2 & 1.0 & 0.9 & 0.53 & 2.46 & 2.96 & 16.5 & 18.5 & 29.8 & 0.82 & 0.84 & 0.91 \\
   \hline
   22 & 12.4 & 21.0 & 13.0 & 1.7 & 0.8 & 0.7 & 0.26 & 1.36 & 1.49 & 5.2 & 7.8 & 13.9 & 0.27 & 0.36 & 0.35 \\
   \hline
   23 & 21.3 & 37.0 & 40.0 & 4.8 & 0.9 & 0.8 & 0.003 & 0.09 & 0.11 & 2.4 & 7.6 & 5.8 & 0.65 & 0.60 & 0.55 \\
   24 & 21.9 & 39.0 & 40.0 & 8.6 & 2.2 & 1.7 & 0.007 & 0.12 & 0.19 & 2.3 & 5.2 & 3.7 & 0.24 & 0.26 & 0.22 \\
   \hline
   Geo.Mean   & 15.0 & 26.9 & 21.6 & 3.2 & 1.1 & 1.0 & 0.15 & 1.34 & 1.44 & 5.5 & 9.5 & 11.1 & 0.32 & 0.40 & 0.39 \\
$\times/\div$ & 1.3 & 1.3 & 1.5 &1.9& 1.7 & 1.6 & 5.5 & 3.98 & 3.63 & 3.3 & 2.9 & 3.0 & 2.46 & 2.36 & 2.49 \\
   \hline
\enddata
\end{deluxetable}

\section{Results}\label{result}

Results for all events are shown in Table~2.  The main salient features are as follows:

\begin{itemize}
\item The loop densities $n_u$ and $n_G$ from each of the dense acceleration region models are typically smaller, by a factor $\sim$3, than the value $n_t$ deduced for the tenuous acceleration region model. This is because in dense models collisional energy loss also occurs within the acceleration region itself, so that the overall density does not have to be as high to accomplish attenuation of the electron flux over the overall distance that is evident from the electron flux images;
\item The sizes of the acceleration region $L_{0,u}$ and $L_{0,G}$ in dense acceleration region models are generally somewhat larger than the acceleration region lengths $L_{0,t}$ deduced from the tenuous acceleration region model.
\item The filling factors $f_u$ and $f_G$ for the dense acceleration region models are about an order of magnitude larger than the values of $f_t$ obtained from the tenuous model. Further, both $f_u$ and $f_G$ are consistent with (the upper limit value of) unity, albeit with a large ($1\sigma$) spread of a factor of about four. For some events (e.g., Events 8, 9, 10, 11 and 15), the values of both $f_u$ and $f_G$ are in excess of unity by an unacceptably large factor.  These events are also characterized by unreasonably large values of the acceleration region length $L_0$ (comparable to, or even larger than, the overall extent of the observed loop) and, in general, the $\chi^2$ values of these fittings are unacceptably large (see Table~2), so that these values of $f$ and $L_0$ should not be taken too seriously.
\item The values of the specific acceleration rate $\eta(\ge$20~keV) for both of the ``dense'' acceleration region models are about a factor of two larger than in the ``tenuous'' acceleration region model.  This is because the neglect of collisional losses in the acceleration region (effectively modeled by assuming a lower density within the acceleration region) implies a smaller number of total particles ${\cal N}$ in the acceleration region, so that a given acceleration rate $d{\cal N}/dt (\ge E_0)$ (deduced from observations of the spatially-integrated hard X-ray flux) corresponds to a larger {\it specific} acceleration rate.  The geometric mean values for $\eta(\ge$20~keV) are  $\sim$5$\times 10^{-3}$~s$^{-1}$ for the tenuous acceleration region model and $\sim$10$\times 10^{-3}$~s$^{-1}$ for the dense acceleration region model (for both uniform and Gaussian spatial injection profiles).  Each of these values has a 1$\sigma$ spread of about a factor of three.
\item The $\chi^2$ goodness-of-fit measures for all three models are comparable.  This indicates that not only is the detailed spatial profile of the electron acceleration poorly constrained by the observations, but also that the decision whether to include collisional modification to the electron spectrum in the acceleration region itself is not driven by the data, but rather by the theoretical considerations related to the physical consistency of the overall scenario.  Given this, it is also possible that a ``hybrid'' model, in which there is a difference (or longitudinal gradient) in the collisional energy loss term (effectively modeled by a spatially varying density) is also a viable model.
\end{itemize}

Note that some of the flares studied are located at significant distances from the solar limb \citep[see Figure~2 in][]{guoetal2012} and consequently the inferred loop lengths may be underestimated due to line-of-sight foreshortening effects.  Further, given that the loop lengths generally increase with energy, the apparent source length $L$, and so volume $V$, at higher energies generally involves a greater portion of the foreshortened parts of the magnetic loop, and thus the foreshortening effect may be more significant at high energies than at lower energies.  According to Equations~(\ref{volume}) through~(\ref{calN}), underestimating the loop lengths $L$ (and hence volumes $V$) will tend to result in overestimation of both the filling factor $f$ and the specific acceleration rate $\eta$.  However, foreshortening effects also reduce the coefficient of $E^2$ in Equation~(\ref{model-electron-tenuous}), causing the inferred loop densities $n_t$ to be overestimated in the tenuous acceleration model; similar results hold for the expressions for $L_U$ and $L_G$ in the dense acceleration model -- Equations~(\ref{model-dense_rec}) and~(\ref{model-dense_gauss}). According to Equations~(\ref{filling-factor}) and~(\ref{calN}), overestimating the density $n$ tends to produce the opposite effect of an increased volume $V$, i.e., {\it smaller} values of both the filling factor $f$ and the specific acceleration rate $\eta$. Thus, overall, these two effects offset, so that the inferred values of the filling factor and specific acceleration rate are relatively insensitive to the effects of foreshortening.  Nevertheless, it must be acknowledged that the effects of foreshortening are a possible source of uncertainty in our results.

\section{Implications for Acceleration Models}\label{discussion}

In this section we discuss the significance of the results obtained, with particular attention to the values of the specific acceleration rate $\eta$.  Specifically, we compare the values with the predictions of candidate electron acceleration models, respectively involving acceleration by large-scale sub-Dreicer electric fields \citep[see, e.g.,][]{1994ApJ...435..469B}, by super-Dreicer electric fields in a reconnecting current sheet \citep{litvinenko1993,litvinenko1996par,litvinenko2000}, and by stochastic acceleration processes.

\subsection{Sub-Dreicer Acceleration}\label{sub_dreicer}

\begin{figure}[pht]
\begin{center}
\includegraphics[width=0.6\textwidth]{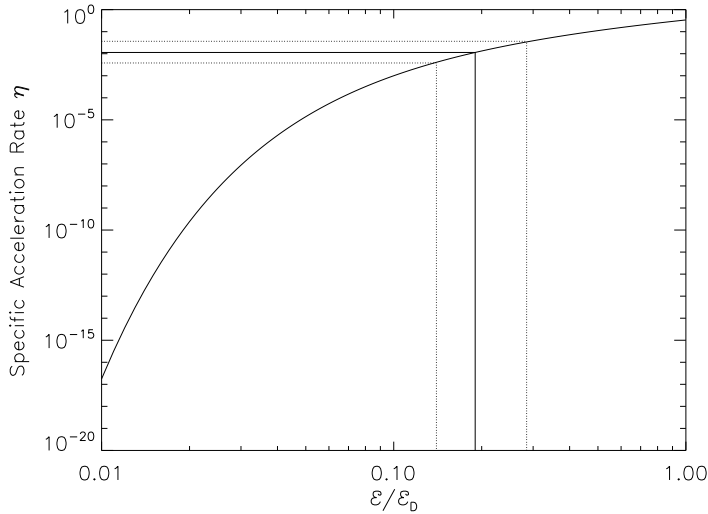}
\caption{The variation of the specific acceleration rate $\eta$ with electric field strength ${\cal E}$ (in units of the Dreicer field ${\cal E}_{\rm D}$). The solid and dashed lines correspond respectively to the geometric mean value $\eta \sim 1.0 \times 10^{-2}$ (electrons~s$^{-1}$ per ambient electron) in the dense acceleration region model and its 1$\sigma$ ranges ($\times/\div$ 3.0) as shown in Table~2.}
\label{fig_Ed}
\end{center}
\end{figure}

Application of a weak, sub-Dreicer, electric field ${\cal E}$ over an extended distance leads to runaway acceleration of those electrons in the high-energy tail of the background thermal distribution for which the applied electric force exceeds the initial Coulomb drag force.  An empirical expression for the specific acceleration rate $\eta$ in such a scenario has been given by \citet{emslie2008AIP} [after \citet{1964PhFl....7..407K}, \citet{1976PhFl...19..239C}, and
\citet{1977PhLA...63..307S}], viz.

\begin{equation}\label{eq:subDreicer}
\eta \sim \left ( 0.3+1.5 {{\cal E} \over {\cal E_{\rm{D}}}} \right ) \left ( {{\cal E} \over {\cal E_{\rm{D}}}} \right )^{-{3\over 8}} \, \exp \left [- \left ( {{\cal E_{\rm{D}}} \over {4 \cal E}} \right ) - \left ( {{2\cal E_{\rm{D}}} \over {\cal E}} \right) ^{1 \over 2 } \right ] \, ,
\end{equation}
where ${\cal E}_{\rm D}$ is the Dreicer field $\simeq 2 \times 10^{-8} \,  n \, $(cm$^{-3}) \, / \, T$~(K) V~cm$^{-1}$.  For the typical densities $n \sim 10^{11}$~cm$^{-3}$ and temperatures
$T \sim$2~keV~$\simeq 2 \times 10^7$~K involved (see Tables~\ref{table:spec} and~2), ${\cal E}_{\rm D} \sim 10^{-4}$~V~cm$^{-1}$. As shown in Figure~\ref{fig_Ed}, the variation of $\eta$ with ${\cal E}$ is very strong; $\eta$ varies by more than fifteen orders of magnitude over the range ${\cal E}/{\cal E}_{\rm D} = 0.01$~to~1.

For the dense acceleration region model, the average value of $\eta$ for the 24 events studied (Table~2) is $\sim 1.0 \times 10^{-2}$, with a multiplicative uncertainty of about 3.  (As noted in Section~\ref{discussion}, these values do not depend critically on the spatial form of the injection profile.)  These bounds on $\eta$, and the corresponding bounds on ${\cal E}/{\cal E}_{\rm D}$, are shown in Figure~\ref{fig_Ed}, from which it is apparent that for the sub-Dreicer model to account for the observed range of $\eta$ values, the applied electric field ${\cal E}$ must be in the very narrow range from about $0.15 \, {\cal E}_{\rm D}$ to $0.3 \, {\cal E}_{\rm D}$, with an average value $\simeq 0.2 \, {\cal E}_{\rm D}$.  Such an electric field strength ${\cal E} \simeq 2 \times 10^{-5}$~V~cm$^{-1}$ rather nicely accelerates electrons up to the threshold energy $\sim$20~keV, but only by using all the available $(10-15)$~arcsecond ($\sim 10^9$ cm) half-length of the coronal loop.  Hence, contrary to claims by \citet{1994ApJ...435..469B}, this mechanism cannot account for the appearance of a power-law spectrum of accelerated electrons up to much higher energies $\gapprox 100$~keV.  Further, it is extremely unlikely that the electric field strength (in units of the Dreicer field) lies within a factor of two in all 24 events studied.

A similarly narrow range of electric field values is obtained if a tenuous acceleration region model is assumed to hold for all events, albeit with a slightly smaller value of ${\cal E}/{\cal E}_D$.  Indeed, even allowing for a variation between tenuous and dense acceleration region models from event to event, the allowable range of ${\cal E}/{\cal E}_D$ values is very small.  We therefore reject the hypothesis that acceleration by large-scale sub-Dreicer fields is responsible for the acceleration of electrons to hard-X-ray-producing energies in the events studied.

\subsection{Super-Dreicer Acceleration}\label{super_dreicer}

In this model \citep{litvinenko1993,litvinenko1996par,litvinenko2000}, the electric field ${\cal E} = (v/c) \times B$ is generated inductively by advective motions in the vicinity of the current sheet associated with the reversal of the principal component of the magnetic field.  For field strengths $B \simeq 100$~G and inflow speeds of order the Alfv\'en speed $v_A = B / \sqrt{4\pi n m_p} \simeq 7 \times 10^7$~cm~s$^{-1}$ (where $m_p = 1.67 \times 10^{-24}$~g is the proton mass), ${\cal E}$ is of order 100~V~cm$^{-1}$, far in excess of the Dreicer field.  Therefore collisional effects in the core acceleration region itself are negligible; after being accelerated for only a few meters, the electrons reach hard-X-ray-producing energies of order 100~keV, and the collisional modification to the accelerated electron spectrum that defines the ``dense acceleration region model'' actually occurs between current sheets in the overall volume occupied by such sheets. Before the accelerated electrons can reach even higher energies, they escape the current sheet because of the presence of a perpendicular magnetic field component that directs the particles out of the sheet. The specific acceleration rate in such a model is given by \citep{emslie2008AIP} [after \citet{1992ASSL..172.....S}]:

\begin{equation}\label{eq:superDreicer}
\eta \sim \left ( {B_\perp \over B_\parallel} \right ) \, \left ({{2 v_A} \over L_s } \right ) \, ,
\end{equation}
where $B_\perp$ and $B_\parallel$ are the perpendicular and parallel components of the magnetic field and $L_s$ is the length of the current sheet, which we take to be comparable to the inferred length $L_0$ of the acceleration region containing the multiple current sheets. Thus, setting $L_s \simeq L_0 \simeq 20$~arcsec $\simeq 1.5 \times 10^9$~cm (Table~2), we find that, numerically,

\begin{equation}\label{eq:superDreicernumbers}
\eta \sim 0.1 \left ( {B_\perp \over B_\parallel} \right ) \, {\rm s}^{-1} \, .
\end{equation}

The observationally-inferred typical value $\eta(\ge$20~keV)$\simeq 10^{-2}$~s$^{-1}$ is thus obtained for a field aspect ratio $B_\perp / B_\parallel \simeq 0.1$.  This is not an unreasonable value; furthermore, unlike for the sub-Dreicer model (in which the value of $\eta$ was extremely sensitive to the value of the applied electric field ${\cal E}$), the value of $\eta$ scales only linearly with this aspect ratio. Consequently, the one-order-of-magnitude range of $\eta$ values corresponds to a very reasonable one-order-of-magnitude range in magnetic field aspect ratio: $0.03 \lapprox B_\perp / B_\parallel \lapprox 0.3$.  We therefore conclude that super-Dreicer acceleration in a current sheet (or, more likely, set of current sheets) is a viable model for explaining the inferred values of the specific acceleration rate $\eta$.  However, there are clearly challenges associated with incorporation of such a current sheet geometry into the overall form of the observed hard X-ray images.  Further, the inference of unit emission filling factors $f$ (Section~\ref{filling}), and the assignment
of a unit acceleration region filling factor $f_{\rm acc}$ (Section~\ref{eta_section}) present considerable geometric challenges for such an acceleration model.  Moreover, as noted in Section~\ref{eta_section}, assigning a value
$f_{\rm acc} \ll 1$ in Equation~(\ref{calN}) would
result in a much smaller value of ${\cal N}$ and so, for a given
value $d{\cal N}/dt (\ge E_0)$, a value of $\eta(\ge E_0)$ that is
too high to be accounted for by Equation~(\ref{eq:superDreicernumbers}) for reasonable values of the magnetic field aspect ratio $(B_\perp/B_\parallel)$.

\subsection{Stochastic Acceleration}\label{stochastic}

Stochastic acceleration models are characterized not through the action of a coherent electric field acting over an extended distance, but rather through multiple interactions of the particle with scattering centers, which may take the form of, for example, regions of enhanced plasma turbulence \citep{2004ApJ...610..550P} or enhanced magnetic field strength \citep{miller1996}.  In general \citep{bian2012}, the acceleration efficiency of a stochastic acceleration model is determined by the momentum diffusion coefficient $D_p = \tau_L {\overline {F^2}}$, where ${\overline {F^2}}$ is the mean square magnitude of the accelerating force and $\tau_L$ is the force correlation time in a Lagrangian frame that comoves with the accelerated particle.

Equation (147) of \citet{bian2012} gives the evolution of the momentum phase-space distribution function $f(p,t)$ under the action of a three-dimensional momentum diffusion process characterized by a momentum diffusion coefficient $D(p)=D_0 \, p^\alpha$:

\begin{equation}\label{f_evol}
f(p,t) \sim t^{3/(2-\alpha)} \, \exp \left [ {-p^{2-\alpha} \over (2-\alpha)^2 \, D_0 \, t} \right ] \, .
\end{equation}
Integrating from reference momentum $p_0$ to $\infty$ shows that the number of particles accelerated above reference momentum $p_0$ within time $t$ is

\begin{equation}
{\cal N}(p_0,t) \sim \int_{p_0}^\infty p^2 \, f(p,t) \, dt \sim
t^{6 / (2-\alpha)} \, \int_{x_0}^\infty e^{-x} \, x^{(1 + \alpha)/(2 - \alpha)} \, dx \, ; \qquad x_0 = {p_0^{2-\alpha} \over {(2-\alpha)}^2 \, D_0 \, t} \, .
\end{equation}
The dependence of ${\cal N}$ on $p_0$ arises from the appearance of
the lower limit $x_0$ is the integral.  Therefore, the characteristic rate (s$^{-1}$) at which particles are accelerated to momenta greater than $p_0$ depends on the magnitude of the diffusion coefficient $D_0$ according to

\begin{equation}\label{eta_stochastic}
\eta \sim (2-\alpha)^2 \, D_0 \, p_0^{\alpha-2} = (2-\alpha)^2 \, {D(p_0) \over p_0^2} \, .
\end{equation}
With $\eta \simeq D_p/p^2 = \tau_L {\overline {F^2}}/p^2$ and writing ${\overline {F^2}} = p^2/\tau_{\rm acc}^2$, where $\tau_{\rm acc}$ is a characteristic acceleration time, we find that the specific
acceleration rate is related to the correlation time $\tau_L$ and the acceleration time $\tau_{\rm acc}$ through

\begin{equation}
\eta \sim {\tau_L \over \tau_{\rm acc}^2} \, .
\end{equation}
Moreover, the total number of scatterings that an electron undergoes can be estimated as the ratio of the acceleration time to the time spent in a single scattering event:

\begin{equation}
N_s \simeq {\tau_{acc} \over \tau_{L}} \simeq {1 \over {\eta \, \tau_{acc}}} \, .
\end{equation}
With observationally-inferred values $\eta \simeq 10^{-2}$~s$^{-1}$, it follows that, if stochastic acceleration is the responsible mechanism,

\begin{equation}
\tau_L \simeq {\tau_{\rm acc}^2 \over 100} ; \qquad N_s \simeq {100 \over \tau_{\rm acc}} \, .
\end{equation}
The acceleration time can be equated roughly with the rise time of the hard X-ray flux, typically a few seconds \citep[see, e.g.,][]{1995ApJ...440..394A}.  Adopting a value $\tau_{\rm acc} = 3$~s, we find that $\tau_L \simeq 0.1$~s and $N_s \simeq 30$.

For stochastic acceleration by cascading magnetohydrodynamic turbulence, \citet[][their Figures 6, 7, 9, 10 and 12]{miller1996}
find a volumetric electron acceleration rate $\sim$(1.5 - 4) $\times 10^8$~cm$^{-3}$~s$^{-1}$ above 20~keV, with the exact value dependent on the parameters and assumptions of the various models considered.  In the \citet{miller1996} model, the background number density is $n=10^{10}$~cm$^{-3}$, so that $\eta \sim$(1.5 - 4) $ \times 10^{-2}$~s$^{-1}$ above 20~keV.  This value of $\eta$ is somewhat higher than the values inferred from our analysis, although it must be noted that the value of the acceleration region density in their model is an order of magnitude lower than that inferred from our observations of extended coronal-loop sources.  We therefore encourage modeling of stochastic acceleration in regions of density $n \sim 10^{11}$~cm$^{-3}$ to determine values of $\eta$ for comparison with our empirically-inferred values.  For deka-keV electrons with velocity $\sim 10^{10}$~cm~s$^{-1}$, the Lagrangian correlation time of order 0.1~s corresponds to a correlation length $\sim 10^9$~cm, comparable to the length of the observed source.

\section{Summary and Conclusions}\label{conclusions}

We have shown that the structure of extended coronal-loop sources can be well explained both by an extended ``tenuous'' acceleration region model, in which the effect of collisions in the acceleration region is ignored (they are assumed either to be overcome by, or to be included in, the acceleration process) and by a ``dense'' acceleration region model, in which post-acceleration collisional energy losses affect the electrons equally inside and outside the acceleration region.  The compatibility of both models with the observations therefore clearly allows for the possibility of a ``hybrid'' model, in which the effect of collisions varies throughout the acceleration and propagation regions, either due to actual density gradients and/or to a spatially-varying dominance of the acceleration process over collisional losses.

Adopting a ``dense'' acceleration region model with a homogeneous collisional loss (density) profile both inside and outside the acceleration region results in the emission filling factor for the hard-X-ray-emitting plasma being closer to unity than in the ``tenuous'' model.  In both models, there are a few anomalous cases, in which unphysical values of $f$ greater than unity are realized.  Discarding these cases, and also adopting a value of unity for the {\it acceleration region} filling factor, the average value of the specific acceleration rate (electrons~s$^{-1}$ per ambient electron) is $\eta(\ge$20~keV) $ \simeq (5-10) \times 10^{-3}$~s$^{-1}$.

Such a value of $\eta$ is difficult to account for in an acceleration models that invokes large-scale acceleration by a weak, sub-Dreicer, electric field.  For such models to be valid, the ratio of the applied electric field ${\cal E}$ to the Dreicer field ${\cal E}_{\rm D}$ must lie in a very close range, extending over less than a factor of two.  Such an acceleration model is, however, consistent with the inferred filling factor $f$ of order unity.

The inferred values of $\eta(\ge$20~keV) are broadly consistent with the predictions of models that invoke large, super-Dreicer, fields formed inductively during the collapse of magnetic field lines in the vicinity of an abrupt field reversal, i.e., a current sheet.  The value of $\eta$ scales linearly with the ``aspect ratio'' of the magnetic field geometry, and it is not unreasonable to expect values of the sheet aspect ratio that extend over the required range from $\sim$(0.03 -- 0.3). However, the issue of the overall geometry of a source composed of multiple thin current sheets remains a significant challenge to this model, particularly in view of the finding that the emission filling factor $f \simeq 1$.

The stochastic acceleration model is consistent with both an emission filling factor $f \simeq 1$ and the values obtained for the specific acceleration $\eta(\ge$20~keV).  However, given the considerable number of degrees of freedom associated with such models \citep[see][]{bian2012}, it must be admitted that achieving agreement with observationally-inferred values is not so much a verification of the model, but rather a way of setting bounds on the parameters that define it.  We have, however, concluded that a viable stochastic model has a Lagrangian correlation time (distance) of order 0.1~s ($10^9$~cm).  Further, during the several seconds of the acceleration process, the accelerated electron is involved in roughly 30 collisions with scattering centers.

There exists very little literature on the predicted value of $\eta$ for stochastic acceleration models.  What little does exist \citep[see, e.g.,][]{miller1996} is based on parameters that are not comparable to the physical environment that characterizes the dense coronal loop sources discussed herein.  We therefore encourage modelers to explore stochastic acceleration models in parameter regimes that are more aligned with the physical conditions in such sources.

\acknowledgements JG and MP have been supported by the EU FP7 Collaborative grant HESPE, grant No. 263086; AGE was supported by NASA Grant NNX10AT78J. JG is partly supported by National NSFC under grant 11233008, by MSTC Program 2011CB811402 and by the German Space Agency (DLR) grant 50 QM 1201. The authors thank Anna Maria Massone, Richard Schwartz, Gabriele Torre, Nicola Pinamonti and Federico Benvenuto for useful discussions, and the referee for pointing out several areas for improvement.

\bibliographystyle{apj}
\bibliography{guo_age}\begin{tiny}

\begin{thebibliography}{26}
\expandafter\ifx\csname natexlab\endcsname\relax\def\natexlab#1{#1}\fi

\bibitem[{{Aschwanden} {et~al.}(1995){Aschwanden}, {Montello}, {Dennis}, \&
  {Benz}}]{1995ApJ...440..394A}
{Aschwanden}, M.~J., {Montello}, M.~L., {Dennis}, B.~R., \& {Benz}, A.~O. 1995,
  \apj, 440, 394

\bibitem[{{Benka} \& {Holman}(1994)}]{1994ApJ...435..469B}
{Benka}, S.~G., \& {Holman}, G.~D. 1994, \apj, 435, 469

\bibitem[{{Bian} {et~al.}(2012){Bian}, {Emslie}, \& {Kontar}}]{bian2012}
{Bian}, N., {Emslie}, A.~G., \& {Kontar}, E.~P. 2012, \apj, 754, 103

\bibitem[{{Brown}(1971)}]{brown71}
{Brown}, J.~C. 1971, Sol.~Phys., 18, 489

\bibitem[{{Cohen}(1976)}]{1976PhFl...19..239C}
{Cohen}, R.~H. 1976, Physics of Fluids, 19, 239

\bibitem[{{Emslie}(1978)}]{1978ApJ...224..241E}
{Emslie}, A.~G. 1978, \apj, 224, 241

\bibitem[{{Emslie} {et~al.}(2008){Emslie}, {Hurford}, {Kontar}, {Massone},
  {Piana}, {Prato}, \& {Xu}}]{emslie2008AIP}
{Emslie}, A.~G., {Hurford}, G.~J., {Kontar}, E.~P., {Massone}, A.~M., {Piana},
  M., {Prato}, M., \& {Xu}, Y. 2008, in American Institute of Physics
  Conference Series, Vol. 1039, American Institute of Physics Conference
  Series, ed. {G.~Li, Q.~Hu, O.~Verkhoglyadova, G.~P.~Zank, R.~P.~Lin, \&
  J.~Luhmann }, 3--10

\bibitem[{{Guo} {et~al.}(2012{\natexlab{a}}){Guo}, {Emslie}, {Kontar},
  {Benvenuto}, {Massone}, \& {Piana}}]{guoetal2012aa}
{Guo}, J., {Emslie}, A.~G., {Kontar}, E.~P., {Benvenuto}, F., {Massone}, A.~M.,
  \& {Piana}, M. 2012{\natexlab{a}}, \aap, 543, A53

\bibitem[{{Guo} {et~al.}(2012{\natexlab{b}}){Guo}, {Emslie}, {Massone}, \&
  {Piana}}]{guoetal2012}
{Guo}, J., {Emslie}, A.~G., {Massone}, A.~M., \& {Piana}, M.
  2012{\natexlab{b}}, \apj, 755, 32

\bibitem[{{Kontar} {et~al.}(2011){Kontar}, {Hannah}, \&
  {Bian}}]{2011ApJ...730L..22K}
{Kontar}, E.~P., {Hannah}, I.~G., \& {Bian}, N.~H. 2011, \apjl, 730, L22

\bibitem[{Krucker {et~al.}(2008)Krucker, Battaglia, Cargill, Fletcher, Hudson,
  MacKinnon, Masuda, Sui, Tomczak, Veronig, {et~al.}}]{krucker08}
Krucker, S., Battaglia, M., Cargill, P., Fletcher, L., Hudson, H., MacKinnon,
  A., Masuda, S., Sui, L., Tomczak, M., Veronig, A., {et~al.} 2008, A\& A
  Review, 16, 155

\bibitem[{{Kruskal} \& {Bernstein}(1964)}]{1964PhFl....7..407K}
{Kruskal}, M.~D., \& {Bernstein}, I.~B. 1964, Physics of Fluids, 7, 407

\bibitem[{Lin {et~al.}(2002)Lin, Dennis, Hurford, Smith, Zehnder, Harvey,
  Curtis, Pankow, Turin, Bester, {et~al.}}]{linetal02}
Lin, R., Dennis, B., Hurford, G., Smith, D., Zehnder, A., Harvey, P., Curtis,
  D., Pankow, D., Turin, P., Bester, M., {et~al.} 2002, Sol. Phys., 210, 3

\bibitem[{Litvinenko(1996)}]{litvinenko1996par}
Litvinenko, Y. 1996, ApJ, 462

\bibitem[{{Litvinenko} \& {Craig}(2000)}]{litvinenko2000}
{Litvinenko}, Y.~E., \& {Craig}, I.~J.~D. 2000, ApJ, 544, 1101

\bibitem[{{Litvinenko} \& {Somov}(1993)}]{litvinenko1993}
{Litvinenko}, Y.~E., \& {Somov}, B.~V. 1993, Sol. Phys., 146, 127

\bibitem[{{Massone} {et~al.}(2009){Massone}, {Emslie}, {Hurford}, {Prato},
  {Kontar}, \& {Piana}}]{massone2009}
{Massone}, A.~M., {Emslie}, A.~G., {Hurford}, G.~J., {Prato}, M., {Kontar},
  E.~P., \& {Piana}, M. 2009, ApJ, 703, 2004

\bibitem[{{Massone} {et~al.}(2004){Massone}, {Emslie}, {Kontar}, {Piana},
  {Prato}, \& {Brown}}]{2004ApJ...613.1233M}
{Massone}, A.~M., {Emslie}, A.~G., {Kontar}, E.~P., {Piana}, M., {Prato}, M.,
  \& {Brown}, J.~C. 2004, \apj, 613, 1233

\bibitem[{{Miller} {et~al.}(1996){Miller}, {Larosa}, \& {Moore}}]{miller1996}
{Miller}, J.~A., {Larosa}, T.~N., \& {Moore}, R.~L. 1996, \apj, 461, 445

\bibitem[{{Petrosian} \& {Liu}(2004)}]{2004ApJ...610..550P}
{Petrosian}, V., \& {Liu}, S. 2004, \apj, 610, 550

\bibitem[{{Piana} {et~al.}(2007){Piana}, {Massone}, {Hurford}, {Prato},
  {Emslie}, {Kontar}, \& {Schwartz}}]{pianaetal07}
{Piana}, M., {Massone}, A.~M., {Hurford}, G.~J., {Prato}, M., {Emslie}, A.~G.,
  {Kontar}, E.~P., \& {Schwartz}, R.~A. 2007, ApJ, 665, 846

\bibitem[{{Singh}(1977)}]{1977PhLA...63..307S}
{Singh}, N. 1977, Physics Letters A, 63, 307

\bibitem[{{Somov}(1992)}]{1992ASSL..172.....S}
{Somov}, B.~V., ed. 1992, Astrophysics and Space Science Library, Vol. 172,
  {Physical processes in solar flares.}

\bibitem[{{Sui} {et~al.}(2004){Sui}, {Holman}, \& {Dennis}}]{suetal04}
{Sui}, L., {Holman}, G.~D., \& {Dennis}, B.~R. 2004, ApJ, 612, 546

\bibitem[{Veronig \& Brown(2004)}]{vebr04}
Veronig, A., \& Brown, J. 2004, ApJL, 603, L117

\bibitem[{{Xu} {et~al.}(2008){Xu}, {Emslie}, \& {Hurford}}]{xuetal08}
{Xu}, Y., {Emslie}, A.~G., \& {Hurford}, G.~J. 2008, ApJ, 673, 576

\end{thebibliography}
•
\end{tiny}

\end{document}